\documentclass[letterpaper,nofootinbib,
preprint,showpacs,preprintnumbers,amsmath,amssymb]{revtex4}%
\usepackage{latexsym}%
\usepackage{hyperref}%

\newcommand{\be}{\begin{equation}}
\newcommand{\ee}{\end{equation}}
\newcommand{\beas}{\begin{eqnarray*}}
\newcommand{\eeas}{\end{eqnarray*}}
\newcommand{\bea}{\begin{eqnarray}}
\newcommand{\eea}{\end{eqnarray}}
\newcommand{\ba}{\begin{align}}
\newcommand{\ea}{\end{align}}
\newcommand{\lb}{\label}

\begin{document}

\title{\bf Hawking radiation from the quantum
           Lema\^{\i}tre--Tolman--Bondi model}
\author{Claus Kiefer}
\email{kiefer@thp.uni-koeln.de}
\affiliation{ Institut f\"ur Theoretische Physik,
Universit\"at zu K\"oln, Z\"ulpicher
Str. 77, 50937 K\"oln, Germany.}
\author{Jakob M\"uller-Hill}
\email{jmh@thp.uni-koeln.de}
\affiliation{ Institut f\"ur Theoretische Physik, Universit\"at zu K\"oln,
Z\"ulpicher Str. 77, 50937 K\"oln, Germany.}
\author{T. P. Singh}
\email{tpsingh@nagaum.tifr.res.in}
\affiliation{Tata Institute of Fundamental Research,
Homi Bhabha Road, Mumbai 400 005, India.}
\author{Cenalo Vaz}
\email{vaz@physics.uc.edu}
\affiliation{RWC and Department of Physics, University of Cincinnati,
Cincinnati, Ohio 45221-0011, USA.}

\begin{abstract}
In an earlier paper, we obtained exact solutions of the Wheeler--DeWitt equation for the 
Lema\^{\i}tre--Tolman--Bondi (LTB)  model of gravitational collapse, employing a lattice 
regularization. In this paper, we derive Hawking radiation in non-marginally bound models 
from our exact solutions.  We show that a non-vanishing energy function does 
not spoil the (approximate) Planck spectrum near the horizon. 
We can also reliably compute corrections 
to the Bogoliubov coefficient because our solutions are exact. The corrections are obtained 
by going beyond the near horizon region and are shown 
to introduce additional greybody 
factors, which modify the black body spectrum of radiation from the black hole.

\end{abstract}

\maketitle


\section{Introduction\label{intro}}

A consistent quantum theory of gravity should provide us with a clear picture of 
black-hole evaporation and black-hole entropy. The main approaches to quantum 
gravity have up to now only led to partial successes \cite{OUP}, concerning 
mainly the statistical explanation for the entropy from microscopic degrees of
freedom. 

In view of the fact that a generally agreed upon quantum theory of gravity 
does not yet exist, it is important to examine the quantization of particular 
models of physical interest by applying established quantization techniques. 
Among the most interesting models are those with spherical symmetry because 
they are sufficiently non-trivial and, at the same time, are simple 
enough to be tractable. One such model is that of Lema\^\i tre, Tolman and Bondi (LTB),
which describes a self-gravitating cloud of non-interacting dust. It was first 
introduced by Lema\^\i tre \cite{lemaitre} in an attempt to describe cosmology, where 
it has found interesting applications \cite{krasinski}. It also describes gravitational 
collapse, but a full understanding of this process requires the application of 
quantum gravity. Given our limited understanding of quantum gravity at present, perhaps 
the approach that is most suited for addressing quantum collapse and questions concerning
its final fate is canonical quantum general relativity. Although limited in 
its scope as a theory of quantum gravity, the canonical theory can meaningfully address
some open questions regarding singularities in the quantum theory, such as the possibility
of singularity avoidance, the quantum evolution of black holes, and the 
role of naked singularities in quantum gravity. 

In  a recent paper we presented an exact canonical quantization of the LTB model \cite{KMV}. 
Earlier work \cite{kuchar,VW99A,VW01,VWS01} had shown that the classical geometrodynamic
constraints of this dust system, the Hamiltonian constraint and the momentum constraint, 
are both given in terms of a canonical chart consisting of the mass contained within spherical 
shells, the physical shell radius, the dust proper time and their conjugate momenta. The 
momentum conjugate to the mass function may 
be eliminated in the Hamiltonian constraint using the momentum constraint, and this procedure 
was shown to lead to a new and simpler constraint which is able to take the place of the 
original Hamiltonian constraint. Dirac's constraint quantization then yields the 
Wheeler--DeWitt equation for the wave functional describing the quantum collapse. 

In \cite{KMV} this approach was sharpened and particular attention was paid to the 
regularization of the Wheeler--DeWitt equation and diffeomorphism invariance. Regularization 
was performed on a one-dimensional lattice 
following a suggestion in \cite{VWS04}, 
and special care was taken to ensure that the momentum constraint is fulfilled 
in the continuum limit. As a consequence, the lattice wave functional becomes described 
in terms of three equations, the Hamilton--Jacobi equation and two additional constraints 
which make it possible to obtain exact solutions
for a particular factor ordering. In this paper we use these exact 
solutions to examine Hawking evaporation during the collapse. The evaporation 
problem was also addressed by us in \cite{VKSW03}, but only for marginally bound collapse in 
the WKB approximation. Here we consider the generic case, that is, the case including the 
non-marginal models for which the classical shells start with a non-vanishing velocity 
at infinity.

Our paper is organized as follows. In Sec.~II we review the major features of the quantum 
LTB model. We present, in particular, the exact quantum states found in \cite{KMV}.  
Sec.~III is the main part of our paper. We show in detail how the spectrum of Hawking 
radiation together with appropriate greybody factors
can be retrieved from these exact quantum states. This provides a bridge from 
quantum gravity (represented here by the LTB model) to semiclassical gravity and the
Hawking effect. Sec.~IV gives a summary of the obtained results and an outlook on potential 
future developments.

\section{The classical and quantum LTB model}

\subsection{The classical LTB model}

The LTB model describes a self-gravitating dust cloud. Its energy--momentum tensor reads
$T_{\mu \nu} = \epsilon(\tau,\rho) u_{\mu} u_{\nu}$, where $u^{\mu}=u^{\mu}(\tau, \rho)$ 
is the four-velocity vector of a dust particle with proper time $\tau$ and labeled by $\rho$
($\rho$ thus labels the various shells that together form the dust cloud). The line element 
for the LTB spacetime is given by
\begin{align}
\label{ltb-metric}
ds^2 &= -d\tau^2 +
\frac{(\partial_{\rho}R)^2}{1+2E(\rho)} d\rho^2
+ R^2(\rho)(d\theta^2 + \sin^2\theta d\phi^2)\ ,
\end{align}
and the Einstein equations lead to
\begin{align}\label{ltb-eg} 8\pi G\epsilon(\tau,\rho) =
\frac{\partial_{\rho}F}{R^2 \partial_{\rho}R}
  \quad \mathrm{and} \quad (\partial_{\tau}R)^2 = \frac{F}{R} + 2E
  \,,\end{align}
where $F(\rho)\equiv 2GM(\rho)$ is some non-negative function. 
We set $c=1$ throughout. 
The case of collapse is described by $\partial_{\tau}R(\tau,
\rho)<0$.

There exists still the freedom to rescale the shell index $\rho$. 
This freedom can be fixed by demanding
\begin{equation} \label{ltb-fix-r} R(0,\rho) = \rho \, , \end{equation}
so that for $\tau=0$ the label coordinate $\rho$ is equal to the curvature radius $R$. 
Now we can express the functions $F(\rho)$ and $E(\rho)$ in terms of the energy density 
$\epsilon$ at $\tau=0$. From \eqref{ltb-eg} one gets
\begin{align} \label{ltb_F} F(\rho) &= 8\pi G
\int_0^{\rho} \epsilon(0,\tilde \rho)\,
\tilde{\rho}^2\,d\tilde \rho\ , \\ E(\rho) &=
[\partial_{\tau}R(\tau=0,\rho)]^2-\frac{8\pi G}{\rho} \int_0^\rho \epsilon (0,\tilde \rho)
\tilde{\rho}^2 \, d\tilde \rho \; .\end{align}
The interpretation of these quantities is that $F(\rho)/2G$ is the active gravitating mass 
inside of $R(\tau,\rho)$, while $E(\rho)$ is the total energy 
per unit mass of the shell labeled by 
$\rho$. The marginally bound models are defined by $E(\rho)=0$. In the present paper we 
discuss the general case which includes the non-marginal case defined by $E(\rho)\neq0$.

The solution of \eqref{ltb-eg} is given in \cite{KMV} and examined in great detail in 
\cite{JS}. The end state of collapse is either a black hole or a naked 
singularity, depending on the initial parameters governing the collapse. At a finite 
dust proper time $\tau=\tau_0(\rho)$, the shell labeled by $\rho$ reaches a curvature 
radius $R=0$, which results in a curvature singularity. Therefore $\tau$ can only 
take values between $-\infty$ and $\tau_0 (\rho)$. 

The canonical formalism begins with the general ansatz 
for a spherically-symmetric line element,
\begin{align}ds^2
  = -N^2 dt^2 + L^2 \left(dr - N^r dt \right)^2 +
  R^2 d \Omega^2  \; , \label{spheric_metric}\end{align}
where $N$ and $N^r$ are the lapse and shift function, respectively. The total action is 
the Einstein-Hilbert action together with an action describing the dust. 
The phase space consists of the three functions $L(t,r)$ and $R(t,r)$ in 
(\ref{spheric_metric}), the dust proper time, $\tau(t,r)$, 
and their canonical momenta $P_L$, $P_R$, and $P_{\tau}$. In the
following, a prime will denote 
a derivative with respect to $r$. All variables are functions of $t$ and $r$. After a 
series of canonical transformations, performed in the spirit of Kucha\v r's 
reduction of static spherical geometries \cite{kuchar}, it becomes possible to describe 
the phase space by a different chart, consisting of the dust proper time, $\tau(r)$, 
the physical radius, $R(r)$, the mass density function, $\Gamma = F'(r)$ and their 
canonical momenta. The collapse problem is reduced to two classical constraints, 
\begin{align}\label{H_new} H &=
  \left({P_{\tau}}^2 + \mathcal{F} \bar{P}_R^2 \right)-\frac{\Gamma^2}{4\mathcal{F}} 
\approx 0 \; ,\\ H_r &= \tau' P_{\tau} + R' {\bar P}_R - \Gamma {P_{\Gamma}}' \approx 0
  \; ,\label{H_new_mc}\end{align}
where $\mathcal{F}=1-F/R$, $\bar{P}_R$ is a new canonical momentum for
$R$, which follows via a canonical transformation from the original
variables, and $P_\Gamma$ is the momentum conjugate to $\Gamma$.
This transformation also absorbs a boundary term 
which is present in the original chart \cite{KMV}.

\subsection{Relation between dust proper time and Killing time}

Under some circumstances, the momentum $P_\Gamma$ has the interpretation of a Killing 
time. In this subsection we explain under which circumstances this interpretation 
makes sense. In the new variables, the proper time is constrained by the expression
\begin{align} \label{lost_tau_prime} \tau' = 2 P_{\Gamma}' \sqrt{1+2E} \pm \frac{R'}
{\mathcal{F}}  \sqrt{1+2E-\mathcal{F}} \; .\end{align}
Defining $a\equiv1/\sqrt{1+2E}$ gives
\begin{align} \tau' = \frac{2P_{\Gamma}'}{a} \pm R' \,
  \frac{\sqrt{1-a^2\mathcal{F}}}{a\mathcal{F}} \; . \label{lost_tau_prime_2}\end{align}
If the mass density vanishes for all $r$ greater than a given $r_b$, and
if $E$ is constant for all $r$ greater than $r_b$, Eq.
\eqref{lost_tau_prime_2} can be integrated. This yields
\begin{align} a \tau &= 2 P_{\Gamma} \pm  \int dR
  \, \frac{\sqrt{1-a^2\mathcal{F}}}{\mathcal{F}} \label{Killing_proper}\\
&= \begin{aligned}[t]2 P_{\Gamma} \pm  F \Bigg(&
  \frac{\sqrt{1-a^2\mathcal{F}}}{1-\mathcal{F}} +
  \ln\left|\frac{1-\sqrt{1-a^2\mathcal{F}}}{1+\sqrt{1-a^2\mathcal{F}}} \right|
  \\ &\quad - \frac{1-a^2/2}{\sqrt{1-a^2}} \, \ln
  \left|\frac{\sqrt{1-a^2\mathcal{F}}-\sqrt{1-a^2}}{\sqrt{1-a^2\mathcal{F}} +
  \sqrt{1-a^2}} \right| \Bigg) \; ,\end{aligned}\label{Ttaurel}\end{align}
where the upper sign refers to infalling dust and the lower sign to outgoing dust. 
Here we have assumed that $0 < a \leq 1$ (corresponding to $E\geq0$). 
The Einstein 
equations guarantee that the discriminant $1-a^2\mathcal{F}$ is
non-negative. (A similar analysis can be carried out for $-1/2< E<0$
\cite{KMV}).

We know from Birkhoff's theorem
that the spacetime around a collapsing dust cloud is given by the Schwarzschild
solution. In \cite{kuchar} it was shown for the Schwarzschild geometry that 
$2P_{\Gamma}$ is equal to the Killing time $T$. Thus 
(\ref{Killing_proper}) 
connects the dust proper time $\tau$ with the Killing time 
$2P_{\tau}$ at the boundary $r_b$. For 
small $\Gamma$ and $E'$ the relationship may still be used, since then we have 
a small amount of dust propagating in a Schwarzschild background; only in this 
case the concept of Killing time makes sense. This is the situation to
be discussed in Sec.~III.
In the limit $a \rightarrow 1$ 
($E\rightarrow 0$) we obtain
\begin{equation} \label{marginaltauT}
\tau = T + 2 \sqrt{F} \left[\sqrt{R} - \frac{\sqrt{F}}{2}
    \ln \left(\frac{\sqrt{R} + \sqrt{F}}{\sqrt{R} - \sqrt{F}} \right) \right] \; ,
\end{equation}
which is identical to the relation used in \cite{VKSW03} for the marginal case. 
The plus sign after the $T$ on the right-hand side has been chosen in order to 
describe a collapsing dust cloud.

The transformation \eqref{Killing_proper} can also be interpreted in
the following way. 
For a Schwarzschild spacetime it gives a relation between 
Killing time and the time used by families of freely falling observers. Each family 
is characterized by a fixed value of $E$. All observers within one family start at 
infinity with the speed $v_{\infty}$, where $E=v_{\infty}^2/(2(1-v_{\infty}^2))$.
In the marginal case they thus start with zero initial speed. In this case  
\eqref{marginaltauT} gives the relation between the Schwarzschild time and the 
Painlev\'e--Gullstrand time \cite{MP}.

\subsection{The quantum LTB model}

Dirac's quantization procedure may be employed to turn the 
classical constraints in \eqref{H_new} and
(\ref{H_new_mc}) into quantum constraints which act on wave functionals. The 
Hamiltonian constraint \eqref{H_new} then leads to the Wheeler--DeWitt
 equation, and the momentum constraint \eqref{H_new_mc} leads to a
 quantum equation which imposes diffeomorphism invariance on the wave
 functional. 

The resulting quantum constraint equations are of a purely formal
nature. They must therefore be regularized before solutions can be
obtained. 
Following a suggestion in \cite{VWS04} we have considered
in \cite{KMV} a one-dimensional lattice given 
by a discrete set of points $r_i$ separated by a distance $\sigma$.
 In order that 
the momentum constraint is fulfilled in the continuum limit, it is important to start
with a corresponding ansatz for the wave functional before putting
it on the lattice. We therefore made the ansatz
\begin{align}\label{ansatz} \Psi\left[\tau(r), \, R(r), \, \Gamma(r)\right] =
U \left( \int dr \, \Gamma(r) \mathcal{W}
(\tau(r), \, R(r),\, \Gamma(r))\right) \; ,\end{align}
where $U: \mathbb{R} \rightarrow\mathbb{C}\,$ is at this stage some
arbitrary (differentiable) function.
The ansatz has to be compatible
with the lattice, which means that it has to factorize into different functions
for each lattice point. So we have to make the choice $U=\exp$, which gives
\begin{align} \lefteqn{\Psi\left[\tau(r), \, R(r), \,
    \Gamma(r)\right]}  \\ &= \lim_{\sigma \rightarrow
  0} \, \prod_i \,\exp \left( \sigma \, \Gamma_i \,
  \mathcal{W}_i\left(\tau(r_i),\, R(r_i), \, F(r_i)\right) \right) \\ &=
\label{factorize_gamma} \lim_{\sigma \rightarrow 0} \,\prod_i \,
\Psi_i\left(\tau(r_i), \, R(r_i), \, \Gamma(r_i), \,  F(r_i)\right) \; ,\end{align}
where
\begin{equation} F(r_i) = \sum_{j=0}^i \, \sigma \,\Gamma_j \; .\end{equation}
The lattice version of the Wheeler--DeWitt equation 
was then found in \cite{KMV} to read
\begin{equation}
 \label{WDW_lattice_gamma}\Bigg[
      G\hbar^2\left(\frac{\partial^2}{\partial \tau_j^2}  + \mathcal{F}_j \,
    \frac{\partial^2}{\partial R_j^2} + A(R_j, F_j)  \,
    \frac{\partial}{\partial R_j}\right) + B(R,F) +
    \frac{\sigma^2 \, \Gamma}{4G \, \mathcal{F}_j}\Bigg] \Psi_j =
    0 \; ,\end{equation}
where $\mathcal{F}_j=1-F_j/R_j>0$ (outside the horizon). Making
the ansatz \eqref{factorize_gamma} together with 
the redefinition $\mathcal{W}=iW/2$, we found in \cite{KMV} that in
order for the resulting equation to be independent of the choice of
$\sigma$ (and thus also in the limit $\sigma\to 0$) a set of {\em
  three} equations must be satisfied. 
If one finds
solutions to all three equations, one can do all other calculations on the lattice,
since these solutions have a well defined continuum limit and satisfy the
momentum constraint. In \cite{KMV} such solutions were found
for a particular factor ordering in the Wheeler--DeWitt equation. 
 (The solutions found in \cite{VWS04} do not
fulfill the momentum constraints in the continuum limit.)

The general solutions within the ansatz (\ref{ansatz}), with $U=\mathrm{exp}$ and 
satisfying the constraints, were found to be of the form
\begin{align}\Psi[\tau, R, \Gamma] =  &\exp(\int dr \, b(F(r))
 \Gamma/G\hbar) \,
  \notag\\ &\times \exp\left\{\mp \frac{i}{2G\hbar} \int dr \,
    \Gamma \;\left[ a(F(r))\tau \pm \int^R dR
    \frac{\sqrt{1-a^2(F(r))\mathcal{F}}}{\mathcal{F}}\right] \right\}
 \; \label{meg_sol}\end{align}
where we have introduced an implicit dependence $a(F(r))$ in order to get solutions 
fulfilling the momentum constraint, and $b(F(r))$ is an arbitrary
function. They serve as the starting point for the discussion of the
Hawking radiation in the next section.


\section{Derivation of Hawking radiation for the general LTB model}

\subsection{Preliminaries}

The exact quantum states \eqref{meg_sol} describe a generic
situation. In order to study Hawking radiation in this framework, we
need to introduce into the formalism the concept of a black hole and
the analogue of the quantum fields used in the standard
treatment. Following \cite{VKSW03}, we consider a Schwarzschild black
hole plus some dust perturbation. The dust perturbation mimics the
quantum fields in Hawking's derivation \cite{Hawking}. Formally, 
this perturbation is
described by the variable $\Gamma\equiv F'$ in which the constant mass
of the black hole drops out. In analogy to \cite{VKSW03}, the
black-hole state factors out in \eqref{meg_sol}, and the remaining
state will describe Hawking radiation. This remaining state then
assumes the form of \eqref{meg_sol} in which $F$ can approximately be
replaced by $2G$ times the black-hole mass.  

The central idea in \cite{Hawking} was the use of a Bogoliubov
transformation of field operators in the Schwarzschild spacetime. Here,
the situation is different, since we have full quantum gravitational
states at our disposal. We thus have to first identify those quantum states
that correspond to the ingoing and outgoing modes, respectively, of
the standard approach. Then we have to calculate their inner
product. Since the description should refer to observers at infinity,
the inner product will be evaluated on hypersurfaces of constant
Killing (Schwarzschild) time $T$, and not on hypersurfaces of constant
dust time $\tau$ (which corresponds to freely falling observers).
 
We would therefore like to re-express our solutions in terms of $T$. For the expanding cloud 
we have the connection between dust time $\tau$ and Killing time $T$ given in (\ref{Ttaurel}):
\be
\lb{N1}
a\tau=T-\int^R dR\ \frac{\sqrt{1-a^2{\mathcal F}}}{\mathcal F}\ .
\ee
The relevant exact solutions of the Wheeler--DeWitt equation read
\be
\lb{N2}
\Psi^{\pm}=\exp\left(\frac{1}{G\hbar}\int dr\ b\Gamma
\mp\frac{i}{2G\hbar}\int dr \ \Gamma\left[a\tau\pm 
\int^R dR\
    \frac{\sqrt{1-a^2{\mathcal F}}}{\mathcal F}\right]\right)\ . 
\ee
In principle one has to include additional normalization factors $N^+$
and $N^-$ for the wave functionals. Since it is not clear how to
normalize the states properly in the full theory, we leave this point
open here; it will not be crucial for our results derived below.

In order to be as close as possible to the standard derivation of Hawking
radiation, we want 
to obtain the scattered outgoing part of the `ingoing wave
functional'. We choose $\Psi^+$ from \eqref{N2} as in-going wave functional 
of positive frequency. This is justified as follows.
Inserting the relation \eqref{N1} into the corresponding
expression in \eqref{N2}, we obtain
\be\lb{in}
\Psi^+=\exp\left(\frac{1}{G\hbar}\int dr\ b\Gamma
 -\frac{i}{2G\hbar}\int dr\ \Gamma T\right)\ ,
\ee
which has the standard form of a positive-frequency wave function.
Note that this wave functional is independent of $R$. 
The lattice version of this state is obtained through the replacement
of $\Gamma=F'$ by a dimensionless quantity, $\omega$. This leads to
\be
\lb{in2}
\Psi^+_{\sigma\omega}=\lim_{\sigma\to 0} \prod_i 
e^{-2b_i\sigma\omega_iF_i/G\hbar-i\sigma\omega_iT_i/G\hbar}\ ,
\ee
see \eqref{factorize_gamma}. Our result will be a product of the
results for each of the shells which together form the dust cloud. 
  
For the out-going modes of negative frequency
 we have to take the state $\Psi^-$. Inserting  
\eqref{N1} into the corresponding state of \eqref{N2} then gives
\be
\Psi^-=\exp\left(\frac{1}{G\hbar}\int dr\ b\Gamma+
\frac{i}{2G\hbar}\int dr\ \Gamma\left[T-2\int^R dR\
    \frac{\sqrt{1-a^2{\mathcal F}}}{\mathcal F}\right]\right)\ .
\ee
The corresponding lattice version reads
\be\lb{out}
\Psi^-_{\sigma\omega}=\lim_{\sigma\to 0}\prod_i
\exp\left(-\frac{2b_i\sigma\omega_i F_i}{G\hbar}-
\frac{i\sigma\omega_i}{G\hbar}\left[ T_i-2\int^R dR_i\
    \frac{\sqrt{1-a_i^2{\mathcal F}_i}}{\mathcal F_i}\right]\right)\ .
\ee
In Sec.~III.B we shall calculate the overlap of in- and out-going
states in the near-horizon limit, while in Sec.~III.C we give a
general expression for the result away from the horizon.

\subsection{The Bogoliubov coefficient in the near-horizon limit.}

We shall now consider the analogue of the usual Bogolubov coefficient
$\beta_{\omega\omega'}$ on each shell. We define it as the inner
product between the in-going wave functional with positive frequency
and the out-going wave functional with negative frequency
(both frequencies are present in the Wheeler--DeWitt equation because
of the $P_{\tau}^2$-term there). In the standard approach one
normalizes the field modes according to the Klein--Gordon inner
product \cite{BD}. For solutions of the Klein--Gordon equation in
$1+1$ dimensions,
\begin{displaymath}
\phi_k(x,t)=u_k(x)e^{-i\omega_kt}\ ,
\end{displaymath}
the spatial modes are normalized according to
\begin{displaymath}
\langle u_k,u_{k'}\rangle=\frac{\delta(k-k')}{2\omega_k}\ ,
\end{displaymath}
from which one gets for the Bogoliubov coefficient
\begin{displaymath}
\beta_{kk'}=2\omega_k\langle u_k^*,\bar{u}_{k'}\rangle\ ,
\end{displaymath}
where $\bar{u}_{k'}$ denotes the `new' set of field modes with respect
to which the solution of the Klein--Gordon equation is expanded.
We thus define in our case
\be
\lb{N6}
\beta_{\omega\omega'} =  
\frac{2\sigma\omega}{G\hbar} \int_{F}^{\infty}dR\
\sqrt{g_{RR}}\Psi^{-*}_{\sigma\omega}\Psi^+_{\sigma\omega'} \ .
\ee
Since we shall evaluate $\beta$ separately for each shell, we skip the
shell index $i$ here and in the following. Note that
$\sigma\omega/G\hbar$ replaces the $\omega$ of the standard approach;
it possesses the correct dimension of an inverse length, so that
$\beta$ itself is dimensionless, as it must be. The integration goes
from the horizon (${\mathcal F}=0$) to infinity. 

In order to specify the measure of integration, we recall again that the
inner product should be evaluated on a $T=$ constant hypersurface
\cite{VKSW03}. Performing the coordinate transformation
$(R,\tau)\mapsto (R,T)$, we get with the help of \eqref{N1} the result
\be
\sqrt{g_{RR}}=\sqrt{1+2E}\frac{R}{R-F}\ .
\ee
We are now ready to evaluate the integral \eqref{N6}. Inserting the
wave functionals 
in \eqref{in2} and \eqref{out} into \eqref{N6} then gives
\bea
& & \beta_{\omega\omega'}= 2\omega \sqrt{1+2E}
\exp\left(-\frac{2b\sigma F(\omega+\omega')}{G\hbar}-i\frac{\sigma
  T(\omega+\omega')}{G\hbar}\right)\times \nonumber\\
& & \int_F^\infty dR \frac{R}{R-F} \exp\left[
\frac{2i\sigma\omega}{G\hbar} 
\int^R dR \frac{\sqrt{1-a^2{\mathcal
        F}}}{\mathcal{F}}\right]\ .
\eea
This is the exact expression, but it is too complicated to compute in the given form. 
However, in order to investigate black-hole radiation, a near-horizon approximation is 
sufficient \cite{VKSW03}. Anyway, we shall derive in the next subsection
an exact, albeit less tractable, result. 
In order to evaluate $\beta$, we introduce the variable
\begin{displaymath}
s=\sqrt{\frac{R}F}-1\ ,
\end{displaymath}
which ranges from zero (horizon) to infinity. The Bogoliubov coefficient, written in
terms of $s$, then becomes
\bea
&&\beta_{\omega\omega'} = 2\omega \sqrt{1+2E}
\exp\left(-\frac{2b\sigma F(\omega+\omega')}{G\hbar}-i\frac{\sigma
  T(\omega+\omega')}{G\hbar}\right)
\times\cr\cr
&& 2F\int_0^\infty ds \frac{(1+s)^3}{s^2+2s}
 \exp\left[\frac{4i \sigma\omega F}{G\hbar} \left(\int^s d\tilde{s} 
(1+\tilde{s})^2 \frac{\sqrt{(1+\tilde{s})^2 -a^2[(1+\tilde{s})^2-1]}}
{\tilde{s}^2+2\tilde{s}}\right)\right].
\lb{exact}
\eea
We have not specified the lower limit in the $\tilde{s}$-integral,
since it will only contribute a phase to $\beta$, which will not
contribute to its absolute square. 

The near-horizon approximation is then described by an expansion with respect to $s$. 
This gives
\bea
\beta_{\omega\omega'} &=& \frac{2F\sigma\omega}{G\hbar}\sqrt{1+2E}
\exp\left(-\frac{2b\sigma F(\omega+\omega')}{G\hbar}-i\frac{\sigma
  T(\omega+\omega')}{G\hbar}\right)
\times \cr\cr
&& \int_0^\infty ds\ s^{-1+2i\sigma\omega F/G\hbar} 
\exp\left[i\frac{\sigma\omega F}{G\hbar} \frac{3+10E}{1+2E}s \right]\ .
\lb{N10}
\eea

In order to evaluate the integral in \eqref{N10}, we insert a regularization factor
$\exp(-ps)$, $p>0$, which guarantees convergence at the upper limit. 
We use the formula \cite{GR}
\begin{displaymath}
\int_0^{\infty}dx\ x^{\nu-1}e^{-(p+iq)x}=\Gamma(\nu)(p^2+q^2)^{-\nu/2}
  e^{-i\nu\mathrm{arctan}(q/p)}
\end{displaymath}
and perform after integration 
the limit $p\to 0$. We then arrive at the following expression: 
\bea
& & \beta_{\omega\omega'}=\frac{2F\sigma\omega}{G\hbar}\sqrt{1+2E}
\exp\left(-\frac{2b\sigma F(\omega+\omega')}{G\hbar}-i\frac{\sigma
  T(\omega+\omega')}{G\hbar}\right)
\times \nonumber\\
& & \ \Gamma\left(\frac{2i\sigma\omega F}{G\hbar}\right)
   \left(\frac{\sigma\omega F}{G\hbar} \frac{3+10E}{1+2E}\right)^{-2i\sigma\omega
F/G\hbar}e^{-\pi\sigma\omega F/G\hbar}\ .
\eea
For the absolute square of $\beta$ we then get
\be
\vert\beta_{\omega\omega'}\vert^2 = \frac{4\pi
  F\sigma\omega(1+2E)}{G\hbar}
\frac{e^{-4b\sigma F(\omega+\omega')/G\hbar}}{e^{4\pi \sigma\omega
    F/G\hbar} - 1}\ .
\ee
In order to calculate the particle-creation rate, we have to evaluate the expression 
$\sum_k\beta_{ik}\beta^*_{ik}$, which here corresponds to the integral
$\int_0^{\infty}d\omega'\ \vert\beta_{\omega\omega'}\vert^2$ (without
further factors, in order to keep the result dimensionless).
We get
\be
\lb{N11}
\langle\mathrm{in}\vert\hat{N}_{\mathrm{out}}\vert\mathrm{in}\rangle=
\frac{\pi\omega(1+2E)}{b}\frac{e^{-4b\sigma F\omega/G\hbar}}
{e^{4\pi \sigma\omega F/G\hbar} - 1}\ .
\ee
where $\hat{N}_{\mathrm{out}}$ denotes the `out'-particle-number operator.

We recognize immediately that the spectrum contains a Planckian
(thermal) factor and that the associated
temperature is independent of $E$, as expected.  
In fact, the only $E$-dependent term is in the prefactor. 

Replacing $\sigma\omega$ by $G\Delta\epsilon$, where 
$\Delta\epsilon$ is the energy of a shell, we arrive at the final result
\be
\lb{result}
\langle\mathrm{in}\vert\hat{N}_{\mathrm{out}}\vert\mathrm{in}\rangle=
\frac{\pi G\Delta\epsilon(1+2E)}{\sigma b}
\frac{e^{-8bGM\Delta\epsilon/\hbar}}{e^{8\pi GM\Delta\epsilon /\hbar} -1}\ ,
\ee
where we have set $F=2GM$. From here we read off the standard 
Hawking temperature for a Schwarzschild black hole:
\be
\lb{temperature}
k_{\mathrm{B}}T_{\mathrm{H}}=\frac{\hbar}{8\pi GM} \ ,
\ee
which holds for each shell separately.

We recognize that the Planckian spectrum is modified by {\em greybody
  factors} which explicitly depend on $\Delta\epsilon$. 
Greybody factors occur in the standard treatment by taking into
account the back-scattering of field modes into the black hole; see,
for example, \cite{Page}. Our greybody factors are different in origin
compared to those coming from back-scattering, since we derived them
from solutions of the Wheeler--DeWitt equation for gravity with
quantum dust. 

At this stage,
the number $b$ has not been fixed further; it occurs in the full
quantum gravitational state and depends on the imposed boundary
conditions. The simplest choice for $b$ is $b=0$. In this case the
integral over $\omega'$ diverges. Dividing out the infinite constant
arising from this integration, one arrives for $b=0$ at the result
\be
\lb{result2}
\langle\mathrm{in}\vert\hat{N}_{\mathrm{out}}\vert\mathrm{in}\rangle=
\frac{8\pi GM\Delta\epsilon(1+2E)}{\hbar}
\frac{1}{e^{8\pi GM\Delta\epsilon /\hbar} -1}\ .
\ee
This corresponds to the result obtained in the WKB approximation, see
\cite{VKSW03} (after taking into account that the $\beta$ there differs
from the $\beta$ here by a factor of $2\omega$).

\subsection{Exact expression for the Bogoliubov coefficient}

It is not difficult to obtain the exact expression for the Bogoliubov coefficient 
by considering the series expansion in $s$ about $s=0$ for arbitrary $E$ of the integrand in 
\eqref{exact}. The measure can be expanded as
\be
\frac{2(1+s)^3}{s^2+2s}=
\frac 1s+\frac 52 + \frac{7s}4+ \ldots \stackrel{\rm{def}}{=} 
\sum_{n=-1}^\infty \alpha_n s^n,
\ee
while the exponent becomes
\bea
&&\frac{4i\sigma\omega F}{G\hbar}\left(\frac 12 \ln s + 
\frac 14(5-2a^2)s+ \frac 1{16}(7-4a^2-2a^4)s^2+ ....
\right)\cr\cr
\stackrel{\rm{def}}{=}
 &&\frac{2i\sigma\omega F}{G\hbar}
\left(\ln s + \sum_{m=1}^\infty \beta_m s^m\right)\ ,
\eea
where we have returned to our notation $a = 1/\sqrt{1+2E}$. 
 From \eqref{exact} we see that we must perform the
 integration 
\be
F\sum_{n=-1}^\infty \alpha_n\int_0^\infty ds\
 s^{n+2i\sigma\omega F/G\hbar}e^{2i\sigma\omega F\beta_1 s/G\hbar}
e^{2i\sigma\omega F\sum_{m=2}^\infty \beta_m s^m/G\hbar}
\ee
where $\beta_1 = \frac 12(5-2a^2) = \frac 12 (3+10E)/(1+2E)$. This is best done by letting
\be
e^{2i\sigma\omega F\sum_{m=2}^\infty \beta_m s^m/G\hbar} = 
1+\frac{2i\sigma\omega F}{G\hbar}\sum_{m=2}^{\infty}\beta_ms^m+
\frac12\left(\frac{2i\sigma\omega
    F}{G\hbar}\right)^2\sum_{m,n=2}^{\infty} \beta_m\beta_ns^ms^n\equiv
\sum_{m=0}^\infty \gamma_m s^m,
\ee
where all the coefficients $\gamma_m$ can be evaluated in principle from the coefficients
$\beta_m$, and $\gamma_0=1$, $\gamma_1=0$. We then find the integral
\bea
&&=F\sum_{n,m=0}^\infty \alpha_{n-1}\gamma_m \int_0^\infty ds s^{-1+n+m+2i\sigma\omega F/G\hbar}
e^{2i\sigma\omega F\beta_1 s/G\hbar}\cr\cr
&&=\left.F\sum_{n,m=0}^\infty \alpha_{n-1}\gamma_m \left(-i\frac{\partial}{\partial \chi}
\right)^{n+m}\int_0^\infty
ds s^{-1+2i\sigma\omega F/G\hbar} e^{i\chi s}\right|_{\chi=2\sigma\omega F\beta_1/G\hbar}.
\eea
In this form, it can be evaluated as before. The first term ($n=0=m$) is the near horizon 
approximation given earlier. The exact Bogoliubov coefficient can be written as the 
(infinite) sum
\bea
\beta_{\omega\omega'} &=&\frac{2F\sigma\omega}{G\hbar}\sqrt{1+2E}
\exp\left(-\frac{2b\sigma F(\omega+\omega')}{G\hbar}-i\frac{\sigma
  T(\omega+\omega')}{G\hbar}\right)
  \times\cr\cr
&&\left(\frac{2\sigma\omega F\beta_1}{G\hbar}\right)^{-2i\sigma\omega F/G\hbar}
e^{-\pi\sigma\omega F/G\hbar}\Gamma\left(\frac{2i\sigma\omega F}{G\hbar}\right)
\times \cr\cr
&& \left[1+ \sum_{n+m=1}^\infty (-i)^{n+m}\alpha_{n-1}\gamma_m
\left(-\frac{2i\sigma\omega F}{G\hbar}\right)\ldots 
\left(-\frac{2i\sigma\omega F}{G\hbar}-n-m+1\right)
\left(\frac{2\sigma\omega F\beta_1}{G\hbar}\right)^{-n-m}\right].\cr
&&
\lb{fullbeta}
\eea
The simplest correction term to the near-horizon approximation is
obtained for $n=1$, $m=0$:
\begin{displaymath}
(-i)\alpha_0\gamma_0\left(-\frac{2i\sigma\omega
    F}{G\hbar}\right)\left(\frac{2\sigma\omega
    F\beta_1}{G\hbar}\right)^{-1}=-\frac{5}{2\beta_1} \ .
\end{displaymath}
Since $1-5/2\beta_1=-2/(3+10E)$, this leads to a
$\omega$-independent correction term in $\vert\beta\vert^2$
given by
\begin{displaymath}
\frac{4}{(3+10E)^2}\leq \frac{4}{9}\ .
\end{displaymath}
In general, the ensuing correction terms 
in \eqref{fullbeta} depend on $\omega$ (and thus on the shell energy
$\Delta\epsilon$) 
and therefore lead to a non-trivial modification of the greybody factors. 
In the limit of $\omega\to\infty$, the correction terms 
are independent of $\omega$.

Such a kind of greybody factors cannot be found in the standard
derivation of Hawking radiation \cite{Hawking}, since there the
geometric-optics approximation is strictly assumed. We can perform
such a calculation because we have an exact quantum state at our
disposal. 

We finally mention that no `trans-Planckian problem' occurs here,
since the whole calculation is based on genuine quantum gravitational
states, and thus all Planck-scale effects are automatically included,
at least within the limits given by our model. 


\section{Summary and Outlook}

The programme for the canonical quantization of the LTB model was started
in \cite{VWS01}, following methods earlier developed in \cite{kuchar}. The 
classical Hamiltonian constraint and momentum constraint were given in terms
of a canonical chart consisting of the mass, radius, dust proper time, and their conjugate momenta. The momentum conjugate to the mass function was eliminated from the Hamiltonian constraint using the momentum constraint, and this procedure
led to a new and simpler Hamiltonian constraint. The Wheeler--DeWitt equation
for the wave functional describing quantum dust collapse then follows from the 
Dirac constraint quantization of the Hamiltonian constraint.

To make progress towards finding solutions to the Wheeler--DeWitt equation, one
must adopt a regularization scheme. The simplest choice is the delta-function
regularization, which was used in \cite{VWS01}, but which turns out to be 
equivalent to the WKB approximation. The solutions for the marginal case in
this approximation were obtained in \cite{VWS01} and were used to derive Hawking
radiation in \cite{VKSW03}. In order to arrive at genuine quantum gravity effects
one clearly needs to go beyond the WKB approximation.

With this intent, a lattice regularization was proposed in \cite{VWS04}. A class of
exact solutions to the Wheeler--DeWitt equation for the general non-marginal case were worked out in \cite{KMV}. In order to construct such solutions, one assumes
that the wave functional of the collapsing dust cloud can be written
as a product of the wave functions of individual shells. We then look
for solutions which are separable in $\tau$ and $R$; it turns out that
for the momentum constraint to be satisfied in the continuum limit, a
specific factor-ordering must be chosen, which happens to depend on
the energy function $E$. Interestingly, the separable exact solutions
which then arise are a complete class of solutions, and there are no
further non-separating solutions. Also, these solutions coincide with
those in the WKB approximation. Thus in \cite{KMV} we found solutions
for which the WKB form is exact -- this is a consequence of the chosen
factor ordering. This implies that the lattice regularization is in
fact equivalent to the WKB approximation, with the additional
constraint that the wave functional of the cloud be a product of shell
wave functionals.

In the present paper we used these exact solutions to derive Hawking radiation
and the accompanying greybody factors for the non-marginal case, thus
generalizing our earlier work for the marginal case in
\cite{VKSW03}. Ideally, one would like to obtain quantum gravitational
corrections to Hawking radiation -- this, however, does not seem
feasible within the framework of the presently used lattice
regularization, since the WKB solution is also an exact solution,
thereby ruling out any scope for a quantum gravity induced
correction. (Interestingly, in \cite{ANOP} it has been suggested that such quantum gravitational corrections to the semiclassical Hawking radiation will be significant only above a certain minimum energy threshold).

Another very important application of a model of quantum gravitational
collapse is to investigate whether or not the singularity predicted
for classical collapse is avoided in the quantum theory. In the
present context this would require us to construct wave packets out of
the exact WKB solutions that we have constructed here. Unfortunately,
that does not seem possible for the non-marginal case since the
factor-ordering we are constrained to choose depends explicitly on the
energy function $E$ (although one could perhaps construct wave packets
with respect to $b$).
On the other hand, it may be possible to
construct wave packets in the marginal case and we plan to investigate singularity avoidance in this case. Also, it would be of interest to investigate the LTB quantum collapse using the methods of loop quantum gravity, although in the loop approach one can at present strictly investigate only the cosmological singularity, and not the black hole singularity arising in an asymptotically flat spacetime.

It also seems useful at this stage to look beyond the lattice regularization,
and investigate other regularization schemes. The Wheeler--DeWitt equation for the LTB model is analogous to a two-dimensional quantum field theory. Hence it will be worth trying to adopt regularization methods used for two-dimensional
quantum field theories to the present gravitational context. We plan to take up
such investigations for the quantum LTB model in the future.
  
\section*{Acknowledgements}

\noindent T. P. S. gratefully acknowledges support from the
German Science Foundation (DFG) under the grant 446 IND 113/34/0-1.




\begin{thebibliography}{99}

\bibitem{OUP} C. Kiefer, {\em Quantum Gravity}. Second edition
 (Oxford University Press, Oxford, 2007).
\bibitem{lemaitre}G. Lema\^{\i}tre, Annales de la Soci\'et\'e
Scientifique de Bruxelles A {\bf 53}, 51 (1933); for an English
translation, see Gen. Rel. Grav. {\bf 29}, 641 (1997).
\bibitem{krasinski} A. Krasi\'nski, {\em Inhomogeneous
Cosmological Models} (Cambridge University Press, Cambridge, 1997).
\bibitem{KMV} C. Kiefer, J. M\"uller-Hill, and C. Vaz, Phys. Rev. D
  {\bf 73}, 044025 (2006).
\bibitem{kuchar} K. V. Kucha\v{r}, Phys. Rev. D {\bf 50}, 3961 (1994).
\bibitem{VW99A} C. Vaz and L. Witten, Phys. Rev. D {\bf 60}, 024009 (1999).
\bibitem{VW01} C. Vaz and L. Witten, Phys. Rev. D {\bf 63}, 024008 (2001);
 {\em ibid.} {\bf 64}, 084005 (2001).
\bibitem{VWS01} C. Vaz, L. Witten, and T. P. Singh,
 Phys. Rev. D {\bf 63}, 104020 (2001).
\bibitem{VWS04} C. Vaz, L. Witten, and T. P. Singh, 
Phys. Rev. D {\bf 69}, 104029 (2004).
\bibitem{VKSW03} C. Vaz, C. Kiefer, T. P. Singh, and L. Witten,
Phys. Rev. D {\bf 67}, 024014 (2003).
\bibitem{JS}P. S. Joshi and T. P. Singh, Phys. Rev. D {\bf 51}, 6778
  (1995); Class. Quantum Grav. {\bf 13}, 559 (1996).
\bibitem{LL} L. D. Landau and E. M. Lifshitz,
{\em The Classical Theory of fields}. Fourth revised English edition
 (Butterworth Heinemann, 1999).
\bibitem{MP} K. Martel and E. Poisson, Am. J. Phys. {\bf 69}, 476
  (2001).
\bibitem{Hawking} S. W. Hawking, Commun. Math. Phys. {\bf 43}, 199
  (1975).
\bibitem{BD}  N. D. Birrell  and P. C. W. Davies, 
           {\em Quantum fields in curved space} (Cambridge University Press,
           Cambridge, 1982).
\bibitem{GR} I. S. Gradshteyn and I. M. Ryzhik, {\em Table of
    Integrals, Series, and Products}, sixth edition (Academic Press,
  San Diego, 2000), p.~343.
\bibitem{Page} D. N. Page, Phys. Rev. D {\bf 13}, 198 (1976);
  A. A. Starobinsky and S. M. Churilov, Sov. Phys. JETP {\bf 38}, 1 (1974).
\bibitem{ANOP} I. Agullo, J.~Navarro-Salas, G. J. Olmo, and L. Parker,
  hep-th/0611355. 

\end{thebibliography}
\end{document}